\documentclass[12pt,colorlinks=true,linkcolor=blue,citecolor=blue,urlcolor=blue]{article}
\makeatletter
\let\frontmatter@title@above=\relax
\makeatother
\usepackage{times}
\usepackage{geometry}
\usepackage[utf8]{inputenc}
\usepackage{enumitem,amssymb}
\usepackage{ragged2e}
\usepackage{setspace}
\usepackage{graphicx}
\usepackage{floatrow}
\usepackage{orcidlink}
\usepackage{makecell}
\usepackage{color}
\usepackage{style_aastex}
\usepackage[para]{threeparttablex}
\usepackage[bf,sf,small,compact]{titlesec}
\titlespacing*{\section}{0pt}{10pt}{6pt}
\titlespacing*{\subsection}{0pt}{3pt}{3pt}
\titlespacing*{\subsubsection}{0pt}{3pt}{3pt}
\newlist{thematic}{itemize}{8}
\setlist[thematic]{label=$\square$}
\usepackage{pifont}

\newcommand{\Rs}{\(\mathsf{R}_\odot\)}

\begin{document}
\pagenumbering{gobble}
\RaggedRight
\noindent {\fontsize{16}{20} \selectfont White Paper for the 2024 Solar \& Space Physics Decadal Survey}
\begin{center}
\title{Radio Studies of the Middle Corona}
{\fontsize{22}{32}\selectfont Radio Studies of the Middle Corona}
\vspace{0.5cm}

\textit{\fontsize{16}{20}\selectfont Current State and New Prospects in the Next Decade}
\end{center}


\normalsize

\justifying


\bigskip

\noindent \textbf{Principal Author:} \\
Bin Chen$^{1}$ \orcidlink{0000-0002-0660-3350} \textit{New Jersey Institute of Technology} \\
Email: \href{mailto:binchen@njit.edu}{binchen@njit.edu}; Phone: (973) 596-3565; Web: \href{https://binchensun.org}{https://binchensun.org}

\smallskip

\noindent \textbf{Co-authors}\\
Jason~E.~Kooi$^{2}$%
\orcidlink{0000-0002-5595-2522}, 
David~B.~Wexler$^{3}$%
\orcidlink{0000-0002-5763-6267}, 
Dale~E.~Gary$^{1}$%
\orcidlink{0000-0003-2520-8396},
Sijie~Yu$^{1}$%
\orcidlink{0000-0003-2872-2614}, 
Surajit~Mondal$^{1}$%
\orcidlink{0000-0002-2325-5298}, \\
Adam~R.~Kobelski$^{4}$%
\orcidlink{0000-0002-4691-1729}, 
Daniel~B.~Seaton$^{5}$%
\orcidlink{0000-0002-0494-2025}, 
Matthew~J.~West$^{5}$%
\orcidlink{0000-0002-0631-2393},
Stephen~M.~White$^{6}$%
\orcidlink{0000-0002-8574-8629}, \\
Gregory~D.~Fleishman$^{1}$%
\orcidlink{0000-0001-5557-2100}, 
Pascal~Saint-Hilaire$^{7}$%
\orcidlink{0000-0002-8283-4556},
Peijin~Zhang$^{8}$%
\orcidlink{0000-0001-6855-5799},
Chris~R.~Gilly$^{9}$%
\orcidlink{0000-0003-0021-9056}, \\
James~P.~Mason$^{10}$%
\orcidlink{0000-0002-3783-5509},
Hamish~Reid$^{11}$%
\orcidlink{0000-0002-6287-3494}

{\fontsize{11}{13}\selectfont \noindent 
[1] New Jersey Institute of Technology; 
[2] Naval Research Laboratory; 
[3] University of Massachusetts, Lowell; 
[4] NASA Marshall Space Flight Center; 
[5] Southwest Research Institute; 
[6] Air Force Research Laboratory;
[7] University of California, Berkeley;
[8] Bulgaria Academy of Sciences, Bulgaria;
[9] Laboratory for Atmospheric and Space Physics;
[10] Johns Hopkins University Applied Physics Laboratory;
[11] University College London
}

\noindent \textbf{Co-Signers \& Affiliations:} see spreadsheet


\vspace{0.2cm}

\noindent \textbf{Synopsis} \\
The ``middle corona,'' defined by \cite{West2022} as the region between $\sim$1.5--6\,\Rs, is a critical transition region that connects the highly structured lower corona to the outer corona where the magnetic field becomes predominantly radial. At radio wavelengths, remote-sensing of the middle corona falls in the meter–decameter wavelength range where a critical transition of radio emission mechanisms occurs. In addition, plasma properties of the middle corona can be probed by trans-coronal radio propagation methods including radio scintillation and Faraday rotation techniques. Together they offer a wealth of diagnostic tools for the middle corona, complementing current and planned missions at other wavelengths. These diagnostics include unique means for detecting and measuring the magnetic field and energetic electrons associated with coronal mass ejections, mapping coronal shocks and electron beam trajectories, as well as constraining the plasma density, magnetic field, and turbulence of the ``young'' solar wind. Following a brief overview of pertinent radio diagnostic methods, this white paper will discuss the current state of radio studies on the middle corona, challenges to obtaining a more comprehensive picture, and recommend an outlook in the next decade. Our specific recommendations for advancing the middle coronal sciences from the radio perspective are:
\begin{itemize}
\item Prioritizing solar-dedicated radio facilities in the $\sim$0.1--1 GHz range with broadband, high-dynamic-range imaging spectropolarimetry capabilities.
\item Developing facilities and techniques to perform multi-perspective, multiple lines-of-sight trans-coronal radio Faraday Rotation measurements.
\end{itemize}

\newpage

\vspace{-2cm}

\section{Introduction}\label{sec:intro}

\pagenumbering{arabic}
\setcounter{page}{1}

The middle corona, defined by \citeauthor{West2022} as the region from $\sim$1.5--6\,\Rs, features a key transition between the inner and the outer solar corona where, in general, the magnetic field of the background corona changes from predominantly closed to radial, and the plasma $\beta$ goes from low to high values. Consequently, this region is highly dynamic in nature and hosts a plethora of key physical processes. While the importance of this region has long been recognized, the middle corona remains poorly understood, due primarily to the lack of observations. The importance of understanding the middle corona and strategic recommendations based on integrated multi-wavelength, multi-messenger observational approaches are discussed in the white paper by \citeauthor{Seaton2022} In this white paper, we focus on the specific aspects that can be addressed by radio observing techniques.

There exist a number of mechanisms relevant to radio emissions from the solar corona, which include gyroresonance (thermal electrons gyrating in the coronal magnetic field), gyrosynchrotron (nonthermal electrons gyrating in the coronal magnetic field), bremsstrahlung (or free-free; electrons interacting with ions), as well as a variety of coherent emissions such as plasma radiation (e.g. the nonlinear growth of Langmuir waves) and electron cyclotron masers (i.e. the nonlinear growth of plasma waves at harmonics of the electron cyclotron frequency). These emission mechanisms co-exist, but because the physical parameters differ in various coronal locations/conditions, the importance of each emission mechanism also varies. In particular, the plasma density $n_e$ and magnetic field $B$ vary dynamically throughout the corona, and hence the corresponding plasma frequency $\nu_p$ and gyrofrequency $\nu_{\rm B}$. As a result, the dominant radio emission mechanism varies over the corona and can change due to local conditions.

\begin{figure}[!ht]
\centering
{\includegraphics[width=0.8\textwidth]{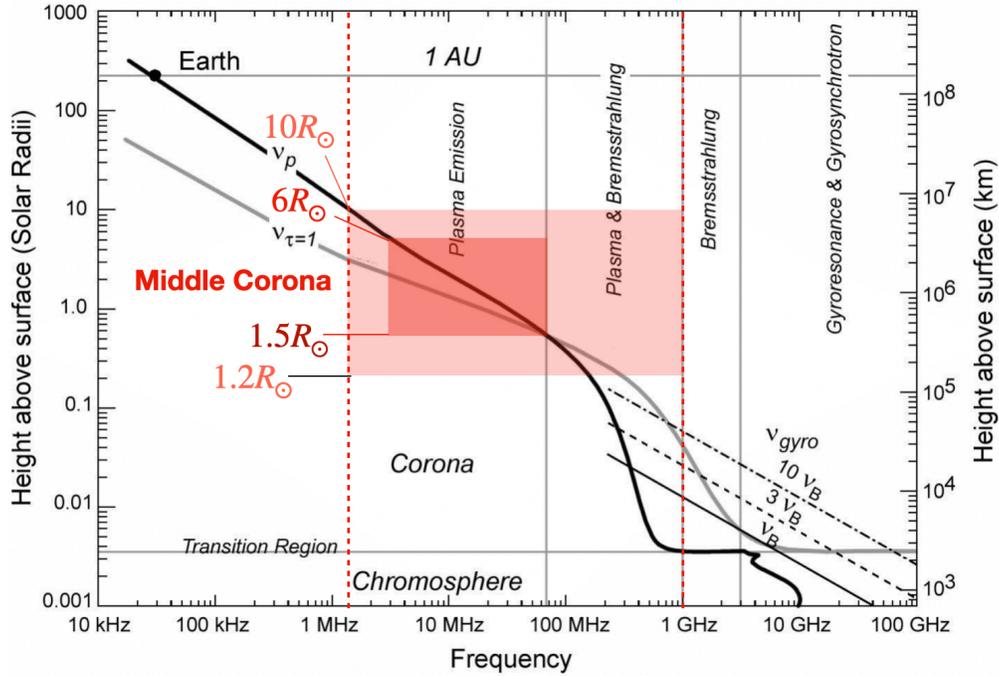}}
\vspace{-0.3cm}
{\caption{Characteristic radio frequencies in the solar atmosphere. The middle corona includes a critical region where the transition of radio emission mechanisms occurs. The dark pink box marks the nominal range of the middle corona ($\sim$1.5--6\,\Rs) and the light pink box marks an extended range taking into account the highly structured and dynamic nature of the corona. The corresponding frequencies that are relevant to radio observations of the middle corona range from $<$10 MHz to $\sim$1 GHz. (Adapted from \citealt{Gary2004} who describe the sources for the temperature, density and magnetic field values used in this plot.)}
    \label{fig:radio_middle_corona}}
\end{figure}

From the perspective of radio observers, the middle corona includes the coronal heights where the mechanism producing solar radio bursts undergoes a key transition from incoherent radio emission to coherent radio emission. Figure~\ref{fig:radio_middle_corona} \citep[adapted from][]{Gary2004} shows the variation of plasma frequency ($\nu_p$; thick black curve), gyrofrequency ($\nu_{\rm B}$; thin black curve), and the frequency of the free-free opacity $\sim$1 ($\nu_\tau=1$) layer as a function of coronal height under typical quiescent coronal conditions. 

The transition region and the innermost inner corona ($\lesssim$ 1.1\,\Rs\ from the center of the Sun) are dominated by incoherent gyromagnetic emission and free-free emission. At around 1.5\,\Rs, the plasma frequency $\nu_p$ layer takes over and becomes higher (closer to the observer) than both the $\nu_\tau=1$ curve and the curves of $\nu_{\rm B}$ and its harmonics. Such a transition has a profound implication for radio observations: the quiescent free-free radio corona is no longer playing a dominant role due to the strong refraction near the plasma frequency. Meanwhile, bright coherent radio bursts, due to plasma radiation occurring near $\nu_{\rm p}$ and its second harmonic (or electron cyclotron maser emission at low harmonics of $\nu_{\rm B}$ if the magnetic field is strong --- see discussion in a white paper by \citeauthor{Yu2022}), start to be important among the observed radio phenomena. Of course, even in the region where the coherent plasma radiation dominates, incoherent radio emission from coronal transients can still be observed, providing crucial diagnostics for these transients. Outstanding examples may include measuring the magnetic field and nonthermal electrons trapped in coronal mass ejection (CMEs) cavities or accelerated by CME-driven shocks (see another white paper by \citealt{Chen2022c}). Therefore, at radio wavelengths, a broad frequency range of $<$10~MHz to $\sim$1~GHz is relevant to the highly dynamic and structured middle corona (light pink box in Figure~\ref{fig:radio_middle_corona}). It is worth emphasizing that the magnetic field and non-thermal electron distribution diagnostics in the middle corona are unique to the radio techniques, since the relatively low density in the middle corona limits emission due to nonthermal particles at other wavelengths (such as X-rays).

Additionally, when radio waves from known, point-like external sources (either natural celestial sources such as quasars/pulsars or spacecraft transmitters) traverse the corona, their propagation and polarization signatures are modulated. The observed temporal and spectral variations provide other means for studying the middle corona. The observed radio signatures can be used to probe the structure and dynamics of the middle corona, including density inhomogeneities, solar wind speed, and coronal magnetic field and its fluctuations. Importantly, these trans-coronal radio sensing methods are applicable in all solar activity states and do not rely on observations of specific episodic outburst phenomena, although spatially resolved diagnostics require measurements along different lines of sight (see white papers by \citealt{Kooi2022a} and \citealt{Kooi2022b}).

\section{Radio Phenomena and Diagnostics in the Middle Corona}
\label{sec:radio_diagnostics}

By virtue of the transitional nature of the radio emission mechanisms in the middle corona, radio observations offer a rich variety of diagnostics for both the background coronal plasma and solar wind, as well as transient phenomena such as CMEs. In the following, we provide a brief overview on various radio phenomena pertinent to the middle corona and the associated diagnostic methods.  

\vspace{-0.5cm}

\begin{figure}[!hb]
\floatbox[{\capbeside\thisfloatsetup{capbesideposition={right,center},capbesidewidth=4.7cm}}]{figure}[\FBwidth]
{\includegraphics[width=0.7\textwidth]{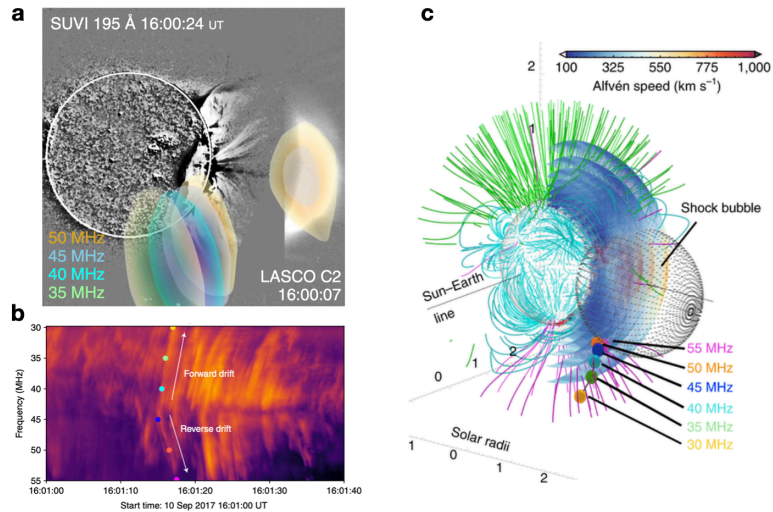}}
{\caption{\fontsize{11}{13}\selectfont LOFAR observations of a decametric type II radio burst event associated with a large-scale coronal shock. a) LOFAR imaging of the burst at the nose and flank of the CME-driven shock. b) ``Herringbone'' fine structures associated with the type II burst, interpreted as the signature of fast electron beams escaping from the shock front. c) 3D model of the magnetic field and the shock geometry. (Adapted from \citealt{Morosan2019}.)}\label{fig:typeII}}
\vspace{-0.5cm}
\end{figure}

\noindent \textit{Type II Bursts and Coronal Shocks} \\
Type II radio bursts are seen from metric to kilometric wavelengths (a few times 100 MHz to 10s of kHz) and are notable for their relatively slow drift to lower frequencies compared to type III radio bursts (see below). They are due to coherent plasma radiation of energetic electrons accelerated at or near the shock front propagating outward at super-Alfv{\'e}nic speeds. Therefore, they bear important diagnostics for both the shock parameters and shock-accelerated electrons. The band-splitting features in type II radio bursts can be used to derive the magnetic Mach number of the shock front \cite{vrvsnak2004band}. Figure~\ref{fig:typeII} shows an example of a type II burst that displays a ``herringbone'' feature in the time-frequency domain. This feature is interpreted as the signature of fast electron beams escaping from the shock front. Recently, thanks to the imaging spectroscopy capability provided by instruments such as LOFAR, new insights have been obtained into their source region at the CME-driven shock (panel (c) of Fig. \ref{fig:typeII}).

\begin{figure}
\floatbox[{\capbeside\thisfloatsetup{capbesideposition={right,center},capbesidewidth=4.5cm}}]{figure}[\FBwidth]
{\includegraphics[width=0.7\textwidth]{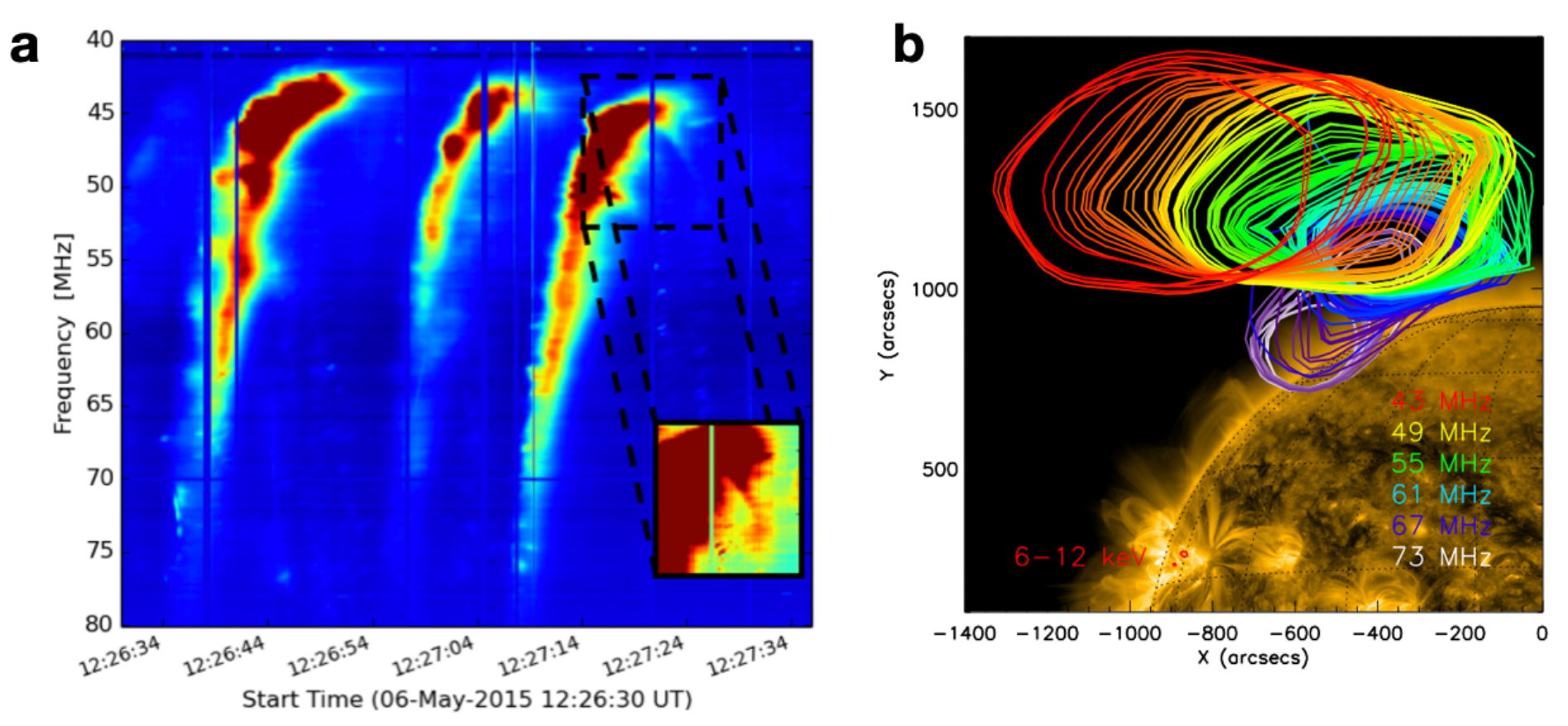}}
{\caption{\fontsize{11}{13}\selectfont LOFAR observations of a type III radio burst event in the middle corona. Multi-frequency images on the right show the trajectory of a type-III-burst-emitting electron beam propagating from low to middle corona. (Adapted from \citealt{Reid2017}.)}\label{fig:typeIII}}
\end{figure}

\begin{figure}[htb!]
\begin{center}
{\includegraphics[width=0.8\textwidth]{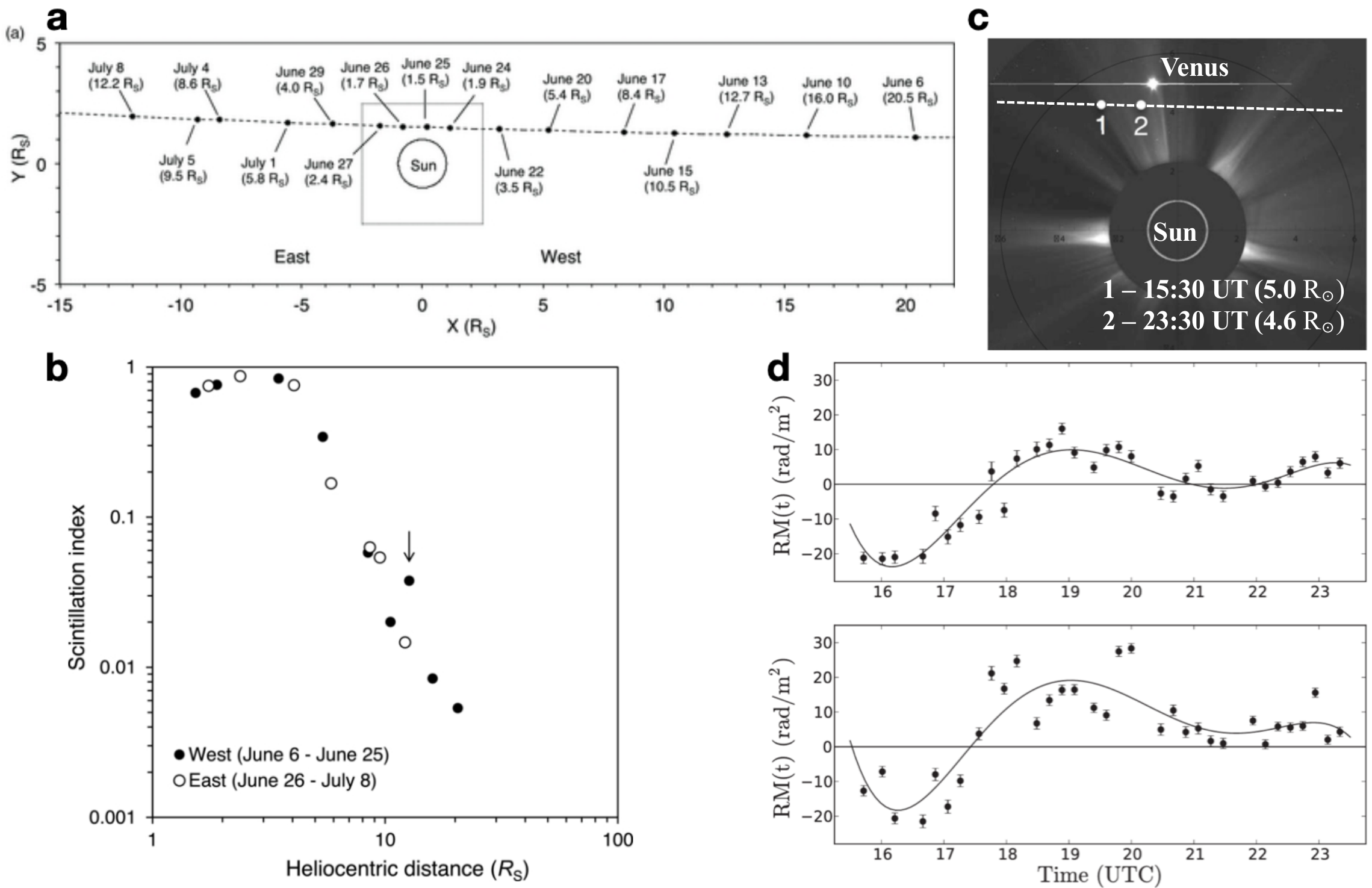}}
\vspace{-0.5cm}
{\caption{\fontsize{11}{13}\selectfont Diagnostics of the middle corona via radio propagation methods. (a) and (b) Radio scintillation index, which represents the magnitude of intensity fluctuations, as a function of heliocentric distance \citep{Imamura2014}. (c) and (d) Faraday rotation provides information on the plasma density and magnetic field component along the line of sight. Differences between measurements along $2+$ lines of sight can be used to probe coronal electric currents \citep{Kooi2014}.} \label{fig:radio_propagation}}
\end{center}
\end{figure}

\noindent \textit{Type III Bursts and Electron Beams} \\
Type III radio bursts are produced by fast electron beams ($\sim$0.1--0.5$c$) escaping along open magnetic field lines (see, e.g., \citealt{reid2014, Reid2020} for recent reviews). Observations of type III bursts span an extremely wide frequency range from $>$GHz to kHz and exhibit a much greater frequency drift than that of type II bursts. In the middle corona, these bursts are predominantly associated with open field lines. With imaging spectroscopy provided by general-purpose facilities such as LOFAR and MWA, new advances have been made in tracing the trajectories of the electron beams (Fig.~\ref{fig:typeIII}) which, in turn, outline the electron-beam-conducting magnetic field lines threading the middle corona \citep[e.g.][]{McCauley2017,Mann2018}. The technique of imaging spectroscopy also has the potential to pinpoint the elusive electron acceleration site(s) \citep{Reid2011}, akin to that achieved in the lower corona with the Jansky VLA \citep{Chen2018}. The emission frequencies and fine structures in the dynamic spectra have been used to derive the coronal density variation in height and properties of the coronal turbulence \citep{Kontar2017,McCauley2018b,Mann2018, Reid2021}.

\smallskip

\noindent \textit{Type IV Bursts and Trapped Electrons} \\
Type IV radio bursts are broadband bursts characterized by their slow- or non-drifting appearance in the dynamic spectrum. Typically observed after the flare peak, they are thought to be produced by nonthermal electrons trapped in closed coronal structures. Depending on the underlying emission mechanism (which can be coherent or incoherent), type IV bursts can, on one hand, outline the closed magnetic structure of interest, and on the other hand, provide diagnostics of the source region \citep[see, e.g., review by][and references therein]{Carley2020b}. Type IV radio bursts have been generally sub-categorized into stationary and moving type IV bursts. The latter, by virtual of their close association with CMEs, are of particular interest.  

\smallskip

\noindent \textit{Radio CMEs} \\
Faint radio emissions that closely resemble their white light CME counterparts are dubbed ``radio'' CMEs because of their similar appearance \citep[see recent review by][]{Vourlidas2020}. 
The emission mechanism is believed to be synchrotron emission, which occurs at large harmonics of the electron gyrofrequency and is thus well above the local plasma frequency, thereby being less affected by the scattering effects (compared to, e.g., the coherent bursts discussed above).  Thanks to their incoherent nature, when imaged at multiple frequencies, radio CMEs can be used to map the coronal magnetic field strength (and, possibly, the plane-of-sky direction if polarimetry is available) and non-thermal electron distribution associated with the CMEs \citep[see, e.g.][]{bastian2001}. For more details on radio diagnostics of CMEs and their progenitors, please refer to another white paper by \citet{Chen2022c}.

\smallskip

\noindent \textit{Propagation Effects} \\
The propagation effects of radio waves provide another means for studying the middle corona. For example, signal broadening and scintillation provide information on the density inhomogeneities in the turbulent coronal plasma \citep[][]{Rickett1990}. Analysis of radio scintillation and frequency fluctuations (Figure~\ref{fig:radio_propagation}(a) and (b)) can provide estimates of solar wind speed \citep{Imamura2014,Wexler2019,Wexler2020}. In addition, modulations of the signal polarization due to Faraday rotation (Figure~\ref{fig:radio_propagation}(c) and (d)) and coronal birefringence can be used to constrain the coronal magnetic field and its fluctuations \citep[see, e.g.][and references therein]{Wexler2017,Wexler2021a,Kooi2021,Kooi2022,kobelski2022}.  

\section{Current Radio Instrumentation}
\label{sec:radio_instrumentation}

Observing the middle corona at radio wavelengths requires accessing a wide frequency range from $<$10 MHz to $\sim$1~GHz (c.f. Figure~\ref{fig:radio_middle_corona}). The $>$20~MHz range is generally accessible from the ground, but the lowest frequencies can only be observed from space due to the ionospheric cutoff. Currently, multiple ground-based instruments are available to observe the frequency range relevant to the middle corona.  In space, new missions, such as the Sun Radio Interferometer Space Experiment (SunRISE; \citealt{Kasper2022}), are being commissioned to locate radio bursts in the upper portion of the middle corona. Table~\ref{tbl:radio_observatories} summarizes the currently operating and upcoming radio facilities that provide imaging capabilities in the frequency range relevant to middle corona studies. This list does not include the large number of radio instruments that provide total-power (full-Sun integrated) dynamic spectral measurements such as the US Air Force's Radio Solar Telescope Network (RSTN) and the e-Callisto network \citep{Benz2009}.

\begin{table}[ht!]
\resizebox{\textwidth}{!}{%
\begin{threeparttable}
\caption{Current radio facilities that provide radio imaging capabilities relevant to middle corona studies.}
\small \label{tbl:radio_observatories}
\begin{tabular}{c c c c c c c}
\hline\hline
Facility & \makecell{Frequency \\ Range} & Elements & \makecell{Max \\ Baseline} & \makecell{Angular \\ Resolution}  & Mid-day UT & \makecell{Solar \\ Dedicated?}\\
\hline\hline
\makecell{MUSER-I\tnote{a} \\ (China)} & 400--2000 MHz & 40 & \makecell{$\sim$3 km} & $\sim$1$'$ at 400 MHz & 04 UT & Yes\\
\hline
\makecell{Nancay\tnote{b} \\ (France)} & 150--450 MHz & 47 & $\sim$3 km  & $\sim$1$'$ at 400 MHz & 12 UT & Yes\\
\hline
\makecell{GRAPH\tnote{c} \\ (India)}  & 40--150 MHz & 32 & \makecell{1.3 km (EW) \\ 0.44 km (NS)} & 5--8$'$ at 150 MHz & 07 UT & Yes\\
\hline
\makecell{OVRO-LWA\tnote{d}} \\ (USA) & 20--88 MHz & 48 & 2.6 km &$\sim$5$'$ at 80 MHz & 20 UT &  Yes\\
\hline
\makecell{uGMRT\tnote{e} \\ (India)} & 120--1450 MHz & 30 & 25 km & 4$''$ at 650 MHz\tnote{j} & 07 UT &  No\\ \hline
\makecell{JVLA P Band\tnote{f} \\ (USA)} & 230--470 MHz & 27 & \makecell{3.4 km \\ (C config.)} & $\lesssim$1$'$ at 400 MHz & 19 UT & No\\
\hline
MWA (Australia)\tnote{g} & 80--300 MHz & 128 &  $\sim$5 km & 1.3$'$ at 150 MHz & 04 UT & No \\
\hline
\makecell{LOFAR \\ (Netherlands)} & \makecell{20--80 MHz (LBA)\tnote{h} \\ 120--180 MHz (HBA)\tnote{i}} & 52 & $>$48 km & $<$16$''$ at 80 MHz\tnote{j} & 13 UT & No\\ 
\hline\hline
\end{tabular}
    \begin{tablenotes}
    \item [a] \citet{Yan2021}.
    \item [b] \citet{Kerdraon1997}.
    \item [c] \citet{Ramesh1998}.
    \item [d] \citet{chhabra2021}.
    \item [e] \citet{gupta2017}.
    \item [f] See \href{https://science.nrao.edu/facilities/vla/docs/manuals/oss/performance/}{JVLA's 2023A performance}.
    \item [g] \citet{beardsley2019}.
    \item [h] \citet{Zhang2022}.
    \item [i] \citet{liu2022interferometric}.
    \item [j] Note the values quoted here are the \textit{theoretical} resolution ($\lambda/d_{\rm max}$). The actual resolution is limited by coronal scattering.
    \end{tablenotes}
\end{threeparttable}}
\end{table}

\begin{figure}[ht!]
\begin{center}
{\includegraphics[width=0.85\textwidth]{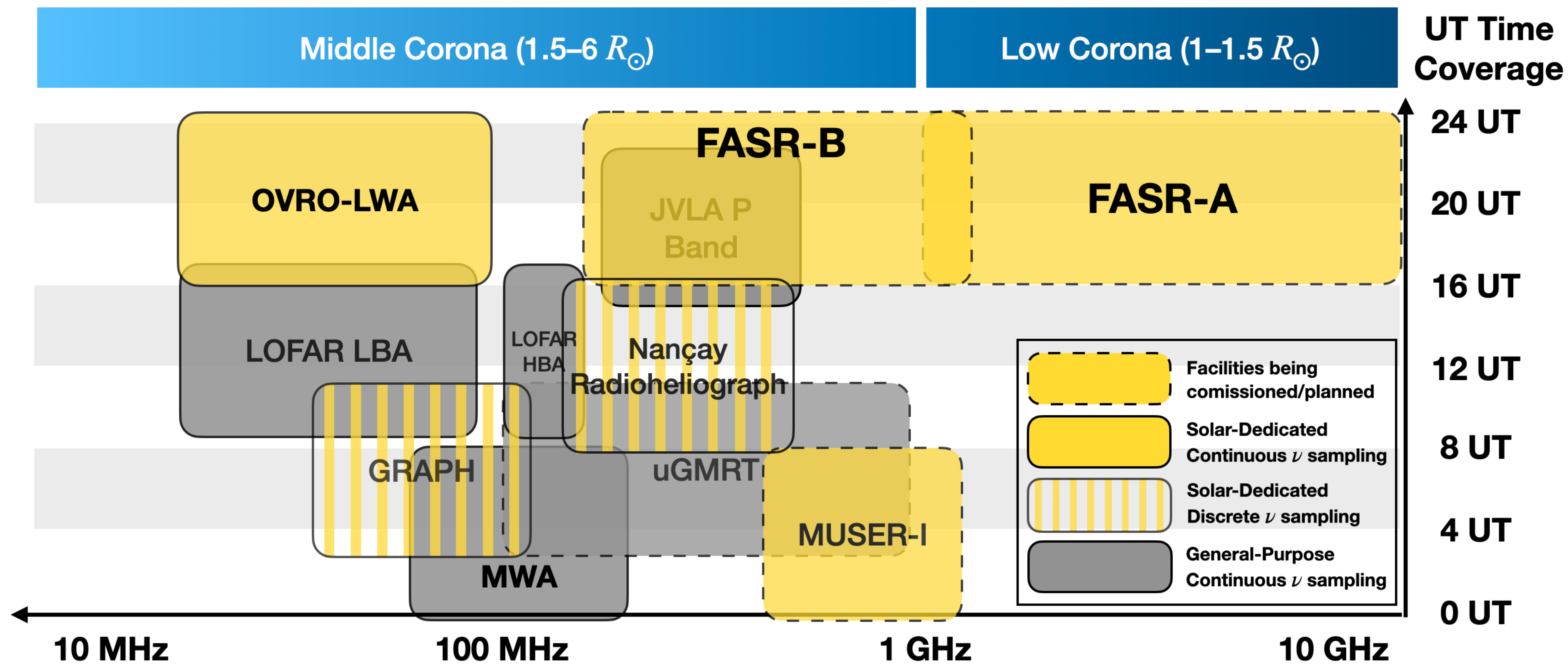}}
{\caption{\fontsize{11}{13}\selectfont Frequency and UT time coverage of radio imaging facilities listed in Table \ref{tbl:radio_observatories}. Filled yellow boxes indicated solar-dedicated instruments with continuous frequency sampling. FASR-B will fill the gap between 0.2--2 GHz in the UT range of $\sim$16--24 UT, by providing radio observations with superior broadband imaging spectropolarimetry. \label{fig:radio_facility}}}
\end{center}
\end{figure}

Over the past decade, new advances have been made with radio facilities equipped with \textit{broadband dynamic imaging spectroscopy}. This exciting new technique allows simultaneous imaging and spectroscopy to be performed over a broad frequency range and at a high time cadence. In other words, a detailed spectrum can be derived from \textit{each pixel} in the radio image for spectral analysis. First realized by the Karl G. Jansky Very Large Array (JVLA) at the decimetric wavelengths \citep{Chen2013} and followed by the commissioning of LOFAR, MWA, EOVSA, and OVRO-LWA, this technique is just beginning to reach the full potential of radio studies using the rich diagnostics tools available \citep[e.g.][]{Carley2020}. Figs. \ref{fig:typeII} and \ref{fig:typeIII} show outstanding examples from LOFAR, which is a general-purpose radio facility (i.e., proposal-driven).

\vspace{-0.5cm}

\section{Future Prospects and Recommendations}

\vspace{-0.2cm}

For radio imaging observations of the middle corona, while multiple facilities are available around the globe, currently \textbf{there is no solar-dedicated radio instrument available to provide true broadband dynamic imaging spectroscopy in the $\sim$0.1--1 GHz spectral range} (see Fig. \ref{fig:radio_facility}). Such a missing capability is critical to producing observations in support of key open questions about the middle corona (see recommendations in \citealt{Seaton2022}). As discussed in Section \ref{sec:radio_diagnostics}, these observations include but are not limited to tracing coronal shocks and shock-accelerated particles, mapping trajectories of fast electron beams from the reconnection site(s), understanding CME initiation and acceleration, understanding CME-accelerated electrons, and providing unique measurements of the evolving magnetic field of CMEs in the middle corona.

To exploit the unique diagnostics of radio observations to the background corona and transient events in the middle corona, the following instrumental requirements must be fulfilled with multiple ground-based and space-borne facilities:
\begin{itemize}
\itemsep0em 
\item Frequency coverage: continuous coverage from $\sim$1 MHz to $\sim$1 GHz to enable access to the entire middle corona. The $\sim$20 MHz--1 GHz range can be done from the ground, while the $\lesssim$20 MHz range needs to be performed from space due to the ionospheric cutoff.
\item High-dynamic-range imaging: a dynamic range of $>$1000:1 is needed at each frequency to detect and map faint and diffuse structures such as radio CMEs. This requirement necessitates a ground-based radio facility with a large number of antenna elements (of order 100).
\item Adequate angular resolution: scattering in the corona limits the usable angular resolution \citep{Bastian1994,Kontar2019}. Resolving structures down to the scattering limit in $\sim$20 MHz--1 GHz translates to an interferometer with a footprint of 3--4 km, which has a resolution of $\sim$1$'$ at 400 MHz. This is sufficient to resolve most features of interest in the middle corona including streamers and CMEs.
\item Adequate spectral resolution: a spectral resolution of $\Delta\nu/\nu\approx 1$\% is needed to observe coherent radio bursts, although an even higher resolution may be necessary to resolve fine structures such as Type III striae \citep[e.g.,][]{Reid2021}.
\item High time resolution: to resolve fast temporal structures of coherent bursts, a time resolution of order 0.1~s is needed. At the high-frequency end ($\sim$1 GHz), a higher time resolution of order 0.01~s is desired \citep[e.g.,][]{Chen2018}. 
\item Dual-polarization performance: Measurements of the total intensity are required as are those of the Stokes V parameter, which contains quantitative information about the magnetic field in, e.g., CMEs.
\end{itemize}


These requirements for advancing radio studies of the middle corona science from the ground already comprise one of the core objectives of the \textit{Frequency Agile Solar Radiotelescope} (FASR) concept, which is envisioned to provide high resolution, high dynamic range, and high fidelity dynamic imaging spectroscopy over a wide frequency range from 0.2--20~GHz. For more details about FASR, we direct the readers to the white paper by \citet{Gary2022c}.

For new frontiers in trans-coronal radio techniques, particularly in extracting the line-of-sight (LOS) magnetic field information of the middle corona in a wide range of helio-altitudes and longitudes, multi-perspective, multiple LOS observations are required. This objective could be achieved by having a large set of small spacecraft fleet transmitting radio signals through the corona at various low impact parameters towards terrestrial radio telescope receivers. In addition, new developments for the Faraday Rotation (FR) tomography method by utilizing 100s of natural background sources can lead to the construction of 2D FR sky maps. A spacecraft fleet concept and new techniques on this front are discussed in white papers by \citet{Kooi2022a} and \citet{Kooi2022b}. 

\bibliography{references}
\bibliographystyle{aasjournal_bc}

\end{document}